\documentclass[pra,twocolumn]{revtex4}
\usepackage{epsfig,amsmath}
\usepackage{subfigure}
\usepackage{graphicx}
\usepackage{dcolumn}
\usepackage{stmaryrd}
\usepackage{mathrsfs}
\usepackage{pifont}
\usepackage{amsthm}
\usepackage{amssymb}
\usepackage{bm}
\usepackage{latexsym}
\usepackage{hyperref}
\usepackage{color}
\usepackage[T1]{fontenc}
\usepackage{float}
\usepackage{amsmath}
\usepackage{amssymb}
\usepackage{stackrel}
\usepackage{graphicx}
\usepackage{esint}
\makeatletter
\pdfpageheight\paperheight
\pdfpagewidth\paperwidth
\floatstyle{ruled}
\newfloat{algorithm}{H}{loa}
\providecommand{\algorithmname}{Algorithm}
\floatname{algorithm}{\protect\algorithmname}

\makeatother

\usepackage{babel}

\begin{document}
	
\title{{Optimally controlled non-adiabatic quantum state transmission in the presence of quantum noise}}
\date{\today}
\author{Xiang-Han Liang$^{1}$, Lian-Ao Wu$^{2,3}$, Zhao-Ming Wang$^{1}$\footnote{%
			wangzhaoming@ouc.edu.cn}}
\affiliation{$^{1}$ College of Physics and Optoelectronic Engineering, Ocean University of China, Qingdao 266100, People's Republic of China \\
$^{2}$ Department of Physics, University of the Basque Country UPV/EHU, 48080 Bilbao, Spain
IKERBASQUE Basque Foundation for Science, 48013 Bilbao, Spain\\
$^{3}$ EHU Quantum Center, University of the Basque Country UPV/EHU, Leioa, Biscay 48940, Spain}

\begin{abstract}
	
{ Pulse controlled non-adiabatic quantum state transmission (QST) was proposed many years ago. However, in practice environmental noise inevitably damages communication quality in the proposal. In this paper, we study the optimally controlled non-adiabatic QST in the presence of quantum noise. By using the Adam algorithm, we find that the optimal pulse sequence can dramatically enhance the transmission fidelity of such an open system. In comparison with the idealized pulse sequence in a closed system,  it is interesting to note that the improvement of the fidelity obtained by the Adam algorithm can even be better for a bath strongly coupled to the system. Furthermore, we find that the Adam algorithm remains powerful for different number of sites and different types of Lindblad operators, showing its universality in performing optimal control of quantum information processing tasks.
}

\end{abstract}

\maketitle

\section{INTRODUCTION}

Information transfer capability lies at the heart of quantum information processing. Likewise, quantum technology requires high fidelity QST through different locations, e.g., between remote microwave cavity memories \cite{axline2018demand}, 
a quantum processor and quantum communication nodes \cite{chapman2016experimental}, matter and light  \cite{matsukevich2004quantum}, from an ion to a photon \cite{stute2013quantum}, from a single photon to a distant quantum-dot electron spin \cite{he2017quantum}, quantum processor and quantum communication nodes \cite{chapman2016experimental}. For short distance communication, quantum spin chain could be a prefect candidate for communication channel \citep{PhysRevLett.92.187902,PhysRevA.85.052319,PhysRevA.83.062328,PhysRevA.91.042321}. Numerous schemes have been suggested to realize perfect or near perfect state transfer \cite{christandl2004perfect,christandl2005perfect,kay2006perfect,zhang2018automatic}. For example, perfect state transfer can be done by construction of the coupling structure of the chain  \cite{christandl2004perfect,christandl2005perfect,kay2006perfect,zhang2018automatic}.
High-fidelity QST through spin chains is made by applying an external field  \cite{fitzsimons2006globally,balachandran2008adiabatic,wang2022nonequilibrium}. 
The high-fidelity QST based on the Floquet-engineered method has been proposed in the many-body problems  \cite{zhou2019floquet}.



Adiabatic evolution has been used in various quantum information processing tasks. QST based on the adiabaticity has also been suggested  for years~\cite{chapman2016experimental, balachandran2008adiabatic,chancellor2012using,chen2018acceleration}. Recently, adiabatic QST in a semiconductor quantum-dot spin chain is studied~\cite{kandel2021adiabatic},
by adiabatically manipulating exchange couplings, and the spin states can be transferred between distant electrons. Typically, adiabatic QST requires a long time. However, the environmental noise will ruin the adiabaticity and this detrimental effect will increase with the evolution time \cite{wang2018adiabatic}.
Consequently, expedited adiabatic processes are desired
\cite{guery2019shortcuts,luan2022shortcuts}. Particular interest in the paper is Ref.~\cite{wang2016shortcut} which suggests to speedup adiabatic processes in terms of various external pulses sequences \cite{wang2018adiabatic}, specifically the acceleration of adiabatic QST in a spin chain under zero-energy-change pulses control \cite{wang2016shortcut}. Adiabatic transmission of an arbitrary entangled state through
an extended SSH chain is also discussed \cite{wang2022arbitrary} , where the topological protection can help to fight against the temporal noise caused by the imperfection in the control field.

For practical quantum state transfer, the corresponding physical communication channel will always suffer from its surrounding environmental noise. The interaction between the system and environment leads to decrease in transmission fidelity between the idealized and the practical \cite{jeske2013excitation,chen2016asymptotically}. For an open system, the environment is Markovian when the memory effects can be neglected. And the Lindblad equation can be used to describe the system dynamics \cite{breuer2002theory,hu2010state}. When the memory effects cannot be neglected, a non-Markovian description is necessary. The non-Markovianity of the environment has a significant influence on the open system \cite{breuer2016colloquium,de2017dynamics}. For example, the memory effects of a non-Markovian environment can be applied to an opto-mechanical system to make it macroscopically entangled  \cite{zhao2019macroscopic}. In general solving the dynamics of the system in a non-Markovian environment is difficult, and the quantum state diffusion (QSD) equation is currently a new-developed method to confront this challenge \cite{ren2019quantum,shi2013non,yu1999non,li2020non,nakajima2018coherent}. The QST through a spin chain between two zero-temperature \cite{ren2019quantum} or finite-temperature non-Markovian baths \cite{wang2021quantum} has been studied using the QSD approach. The transmission fidelity decreases with the strength of the system-bath coupling and temperature. 

On the other hand, besides these previous adiabatic QST proposals, Ref.~\cite{wang2016shortcut} suggests non-adiabatic QST in terms of external pulses in a spin chain. It is interesting to note that the external pulses can somehow wash out the quasi-crossing between different energy levels during the change of the time-dependent Hamiltonian.  In this paper, we will extend the protocol used in \cite{wang2016shortcut} to the zero-energy-cost control pulses and optimize control sequences in the presence of noise.  We will use the QSD equation to investigate the non-adiabatic transport of quantum state in a one-dimensional spin chain in the presence of noise. Zero-energy-cost control has been introduced theoretically \cite{wang2020almost} to realize almost exact state transmission in a spin chain in an open system. For the numerical optimization, we also study the stochastic learning control of adiabatic speedup in a non-Markovian open qutrit system\cite{xie2022stochastic}. The stochastic search procedures are proved to be powerful tools to design control pulses for combating the detrimental environment. We will compare the theoretical and numerical results for the optimal pulses control in the realization of non-adiabatic QST proposals. Specifically, we will check the non-Markovian effects of the environment on the state transmission fidelity.

\section{The model and the Hamiltonian}

The total Hamiltonian of the open quantum system can be written as
\begin{equation}
	H_{tot}=H_{s}+H_{b}+H_{int},
\end{equation}
where $H_s$ and $H_b$ are the Hamiltonian of the system and bath, respectively. $H_{int}$ is the system-bath interaction Hamiltonian. For a bosonic environment, $H_{b}=\Sigma_k\omega_{k}b_{k}^{\dagger}b_{k}$ (for convenience, setting $\hbar=1$), where $b_{k}^{\dagger}$ $(b_{k})$ is the bosonic creation (annihilation) operator of $k$th mode with frequency $\omega_{k}$. The interaction Hamiltonian $H_{int}$ reads
\begin{equation}
	H_{int}=\underset{k}{\sum}(g_{k}^{*}L^{\dagger}b_{k}+g_{k}Lb_{k}^{\dagger}),
\end{equation}
where $g_{k}$ is the coupling strength between the system and the $k$th mode of the bath. $L$ is the Lindblad operator.

Initially, suppose that the bath is prepared at the thermal equilibrium state, the density operator is $\rho(0)=e^{-\beta H_{b}}/Z$ with temperature $T_{em}$. $Z=Tr[e^{-\beta H_{b}}]$ is the partition function and $\beta=1/T_{em}$ (setting $K_{B}=1$). According to Refs.  \cite{wang2022nonequilibrium,nie2021control}, the non-Markovian master equation of the system can be derived by the non-Markovian QSD equation technique \cite{diosi1997non,diosi1998non}.
\[
\frac{\partial}{\partial t}\rho_{s}=-i[H_{s},\rho_{s}]+[L,\rho_{s}\overline{O}_{z}^{\dagger}(t)]-[L^{\dagger},\overline{O}_{z}(t)\rho_{s}]
\]
\begin{equation}
[L^{\dagger},\rho_{s}\overline{O}_{\omega}^{\dagger}(t)]-[L,\overline{O}_{\omega}(t)\rho_{s}],
\label{eq.3}
\end{equation}
where $\overline{O}_{z,(\omega)}=\intop_{0}^{t}ds\alpha_{z,(\omega)}(t-s)O_{z}$, and $\alpha_{z,(\omega)}(t-s)$ is the bath correlation function. Note that in above equation, the weak coupling is assumed and the $O$ operators are assumed to be independent of noises \cite{xu2014perturbation}.

For the bath we choose the Lorentz spectrum, with the spectral density $J(\omega)=\frac{\Gamma}{\pi}\frac{\omega}{1+(\frac{\omega}{\gamma})^{2}}$, where $\Gamma$ and $\gamma$ are real parameters. $\gamma$ represents the characteristic frequency of the bath, and $\Gamma$ represents the strength of the system-bath coupling. With the Lorentz spectrum, the bath correlation functions can be written as
\begin{equation}
\alpha_{z}(t-s)=\Gamma T_{em}\varLambda(t,s)+i\Gamma\varLambda(t,s),
\label{eq.1}
\end{equation}
\begin{equation}
\alpha_{\omega}(t-s)=\Gamma T_{em}\varLambda(t,s),
\label{eq.2}
\end{equation}
where $\varLambda(t,s)=\frac{\gamma}{2}e^{-\gamma|t-s|}$ is an Ornstein-Uhlenbeck correlation function. $1/\gamma$ represents the memory time of the environment. For equations (\ref{eq.1}) and (\ref{eq.2}), we have the relations,
 \begin{equation}
 \frac{\partial \alpha _{z(\omega )}(t-s)}{\partial t}=-\gamma \alpha _{z(\omega )}(t-s).
\end{equation}

And the $\overline{O}_{z,(\omega)}$ operator satisfies  \cite{yu2004non,diosi1998non},
\begin{equation}
\frac{\partial\overline{O}_{z}}{\partial t}=(\frac{\Gamma T_{em}\gamma}{2}-\frac{i\Gamma\gamma^{2}}{2})L-\gamma\overline{O}_{z}
+[-iH_{s}-(L^{\dagger}\overline{O}_{z}+L\overline{O}_{\omega}),\overline{O}_{z}],
\end{equation}
\begin{equation}
\frac{\partial\overline{O}_{\omega}}{\partial t}=\frac{\Gamma T_{em}\gamma}{2}L^{\dagger}-\gamma\overline{O}_{\omega}
+[-iH_{s}-(L^{\dagger}\overline{O}_{z}+L\overline{O}_{\omega}),\overline{O}_{\omega}].
\end{equation}
In the Markov limit, Eq.(\ref{eq.1}) and Eq.(\ref{eq.2}) become $\alpha_{z}(t-s)=\alpha_{\omega }(t-s)=\Gamma T_{em}\delta (t-s)$. $\overline{O}_{z}=\frac{\Gamma T_{em}}{2}L$, and $\overline{O}_{\omega }=\frac{\Gamma T_{em}}{2}L^{\dagger }$. The master equation in Eq. (\ref{eq.3}) reduces to the Lindblad form \cite{diosi1997non,yu1999non,wang2021quantum,yu2004non},
\[
\frac{\partial}{\partial t}\rho_{s}=-i[H_{s},\rho_{s}]+\frac{\Gamma T_{em}}{2}[(2L\rho _{s}L^{\dagger}-L^{\dagger }L\rho _{s}-\rho _{s}L^{\dagger }L)
\]
\begin{equation}
+(2L^{\dagger}\rho _{s}L-L L^{\dagger }\rho _{s}-\rho _{s}L L^{\dagger })].
\end{equation}

For the system Hamiltonian, in this paper, we choose the time-dependent one-dimensional spin chain model as in Ref. \cite{wang2016shortcut},
\begin{equation}
	H_{s}(t)=A(t)H_{xy}+B(t)H_{z},
\end{equation}
where $H_{xy}=J\sum_{i=1}^{N-1}(\sigma_{i}^x \sigma_{i+1}^x + \sigma_{i}^y \sigma_{i+1}^y)$ is the hopping term, and $H_{z}=\sum_{i=1}^{N}h(i)\sigma_{i}^z$ is the on-site energy term. $J$ represents the coupling between the nearest two sites, and now we set $J=-1.0$ throughout.  $N$ is the number of sites.
$\sigma_{i}^x, \sigma_{i}^y, \sigma_{i}^z$ are the Pauli matrices acting on spin $i$. $i$ is the location of the sites, $i=1,2,\ldots,N$. $h(i)$ represents a non-zero gradient field along the $z$-direction of the spin chain  \cite{wang2016shortcut,PhysRevA.88.022323}. For $h(i)$, $h(i)<h(i+1)$. $T$ is the total evolution time. For $A(t)$ and $B(t)$, they satisfy the condition $A(0)=A(T)=0$ and $B(0)=1$, $B(T)=-1$. For simplicity, let $h(i)=h_{m}i$. This model can be realized in an optics lattice \cite{feng2019photonic,wu2004overcoming}. In this case, $A(t)=sin(\varOmega t)$, $B(t)=cos(\varOmega t)$, and $\varOmega=\pi/T$. The model physically describes ultracold atoms in an one-dimensional optical lattice modulated by laser beam  \cite{wang2013fault}. 

Now suppose the initial state of the system is prepared as $|\Phi _s(0)\rangle=|1\cdots 00\rangle$. The target is to transfer the state $|1\rangle$ at the first site to the other end of the chain at some time $T$ with $|\Phi _s(T)\rangle=|0\cdots 01\rangle$. The fidelity can be defined as
\begin{equation}
	F(t)=\sqrt{\langle\Phi_s(T)|\rho_s(t)|\Phi_s(T)\rangle},
	\label{eq12}
\end{equation}
where $\rho_s(t)$ is the system's reduced density matrix.

\section{Quantum state transfer under control}

Normally the existence of the environmental noise will destroy the quantum information processing tasks, e.g., decreasing the state transmission fidelity \cite{wang2021quantum} or adiabaticity \cite{xie2022stochastic}. Quantum control has been applied to resist the detrimental effects of the environment. A recursive method \cite{mouloudakis2022arbitrary} has been used to calculate the state transmission fidelity of arbitrary-length X-X spin chains boundary-driven by non-Markovian environments. Quantum optimal control \cite{wang2021optimal} by adding an leakage elimination operator Hamiltonian to the system \cite{xie2022stochastic} has been suggested to realize adiabatic speedup in a non-Markovian open qutrit system. The leakage elimination operator Hamiltonian can be realized by a sequence of pulses, which can be constructed as \cite{wang2014fast,wang2018adiabatic,wu2022adiabatic}
\begin{equation}
	H(t)=[1+c(t)]H_{0}(t),\label{eq:3}
\end{equation}
where $c(t)$ represents the control function. In Ref. \cite{wang2016shortcut}, $c(t)$ is chosen to be an arbitrary function but always positive, as a result the system average energy will be increased. In this paper, the pulse is chosen as a zero area pulse as in Ref. \cite{wang2022nonequilibrium,xie2022stochastic} with a positive in the first half period and a negative in the second half period. In this case, the average energy of the system is not increased. And the type of the pulses has little influence on the fidelity \cite{zhang2019adiabatic}, so we take the rectangular pulse as an example  \cite{PhysRevA.89.032110,wang2016ultrafast}:
\begin{equation}
	c(t)= \begin{cases}\begin{array}{cccc}
		I,2n\tau<t<(2n+1)\tau,\begin{array}{ccc}
		\end{array}\\
		-I,(2n+1)\tau<t<(2n+2)\tau,
	\end{array}\end{cases}
\end{equation}
where $n=0,1,2,\ldots$, $I$ is the pulse strength and $\tau$ is half period of the pulses. For this kind of pulses, it has been theoretically derived that if the pulses satisfy the condition $I\tau=2\pi m,\;m=1,2,3,\ldots$ \cite{wang2018adiabatic,wu2022adiabatic,he2021adiabatic}, the adiabatic speedup can be realized. Note that this condition is only valid for a fixed energy gap and closed system. In our model, the energy gap $\triangle E_{01}$ between the ground state and the first excited state is time-dependent. 
In this case, the pulse strength can be tuned as $I(t)=I / \triangle E_{01}$ \cite{he2021adiabatic,ren2020accelerated}. In our model, the energy level crossing occurs at $t=T/2$, which leads to an infinite pulse intensity at that point. We then use a suitable value instead at the crossing point.
The Lindblad operation $L=\Sigma_{i}\sigma _i^-$ is used if not especially specified. $\sigma_i^-=\sigma_i^x-i\sigma_i^y$ is the spin lowering operator.

\begin{figure}
	\includegraphics[scale=0.35]{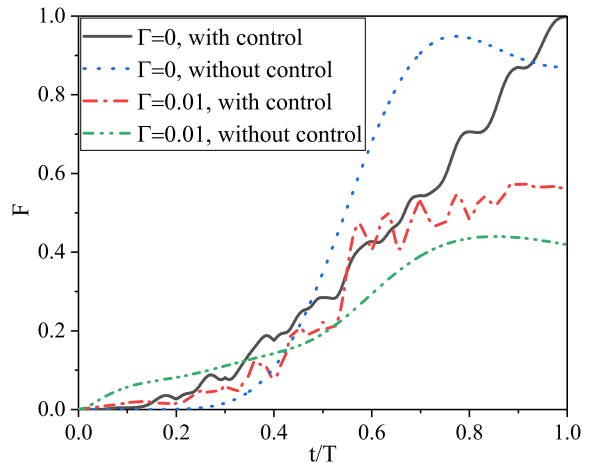}
	\caption{The fidelity $F$ versus the rescaled time $t/T$. The total evolution time $T=\pi$. With control the parameters $I=20, \tau=\pi/10$. In the presence of environment, the parameters are taken as $\Gamma=0.01,\gamma=20,Tem=20$. The number of sites $N=5$.}
	\label{fig:1}
\end{figure}

In Fig.~\ref{fig:1}, we plot the the fidelity as a function of the rescaled time $t/T$ with and without control. For the control, the ideal pulse condition $I\tau=2\pi$ is used with $I=20$, $\tau=\pi/10$. If we do not consider the environment ($\Gamma=0$), the ideal pulse control can be used to dramatically improve the fidelity. Near perfect QST ($F=0.99837$) can be obtained at $t/T=1$. However, when considering the environment ($\Gamma=0.01$), the fidelity is low even under control ($F=0.56041$).

We have stressed that the ideal pulse conditions are derived for a closed system. Now from Fig.~\ref{fig:1} we see that it loses its effectiveness in an open system. Stochastic learning control of adiabatic speedup in a non-Markovian open qutrit system has been studied in Ref.~\cite{xie2022stochastic}. The the stochastic search procedures are proved to be powerful tools for the 
design of control pulses in an open system. Here we will use the Adam algorithm \cite{kingma2014adam}, which is the extended version of stochastic gradient descent, to design the optimal pulse for high fidelity state transmission in the presence of environment.

Now the optimization objective can be denoted as minimizing the loss function, or fidelity error. It is usually defined as
\begin{equation}
	Loss(I^{N})=1-F(I^{N})+\lambda c_{max}.
	\label{eq:14}
\end{equation}
Here $c_{max}$ is the maximum value of the control function $c(t)$. $\lambda$ is a constant, in this paper we choose $\lambda=0.01$. We introduce this term to constrain the control pulse. Eq.(\ref{eq:14}) allows for the competition between the infidelity $1-F(I^{N})$ and the maximum applied control intensity $c_{max}$, thus avoiding the generation of an optimized pulse with too large intensity.

The Adam algorithm can be denoted as follows.

\begin{figure}
	{\includegraphics[scale=0.35]{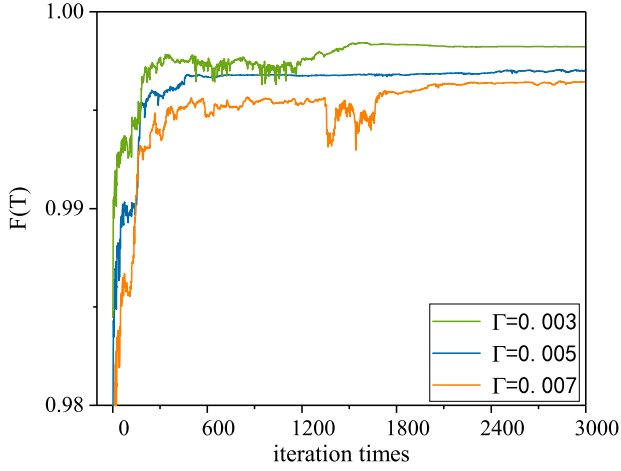}}
	\caption{The variation of the fidelity $F(T)$ vs the number of iteration times for different $\Gamma$. $\gamma=2$ and $Tem=10$.}
	\label{fig:2}
\end{figure}

\textit{Step 1.} Calculate the gradient vector $\boldsymbol{g}$ of the loss function $Loss$ with respect to the selected variable $\boldsymbol{I}$
\begin{equation}
	\boldsymbol{g}=\triangledown_{I} Loss(\boldsymbol{I}).
\end{equation}
\textit{Step 2.} Calculate the new exponential moving average
\begin{equation}
	\boldsymbol{m}=\beta_1 \boldsymbol{m}+(1-\beta_1)\boldsymbol{g}.
\end{equation}
\begin{equation}
	\boldsymbol{v}=\beta_2 \boldsymbol{v}+(1-\beta_2)(\boldsymbol{g})^2.
\end{equation}
\textit{Step 3.} Compute the new bias-corrected moment vectors
\begin{equation}
	\boldsymbol{\hat{m}}=\boldsymbol{m}/(1-\beta _1).
\end{equation}
\begin{equation}
	\boldsymbol{\hat{v}}=\boldsymbol{v}/(1-\beta _2).
\end{equation}
\textit{Step 4.} Update the variables $\boldsymbol{I}$ according to
\begin{equation}
	\boldsymbol{I}=\boldsymbol{I}-\alpha \boldsymbol{\hat{m}}/(\sqrt{\boldsymbol{\hat{v}}}+\varepsilon).
\end{equation}
\textit{Step 5.} Repeat the above steps until $Loss<\xi$ or $k>k_{max}$ ($\xi$ and $k_{max}$ denote the given threshold and the maximum number of iterations, respectively).

For the Adam algorithm, $\boldsymbol{I}$ indicates pulses intensity, $\boldsymbol{g}$ is the gradient. $\beta$ is a fixed parameter. $\alpha$ is the learning rate, $\varepsilon$ is a constant set to avoid the denominator being zero.
$\xi$ is the the given threshold.

The complete algorithm description of Adam is as follows.

\begin{algorithm}
	\caption{Adam}
	\label{algorithm}
	
	Initial pulse intensity $I^{i}$, final pulse intensity $I^{i}$.\\
	\textbf{Parameter:} EMA parameters $\beta_1$ and $\beta_2$, learning rate $\alpha$ and the epsilon $\varepsilon$.\\
	\textbf{for} iteration $k=1,k_{max},m^i=0,\upsilon^i=0$.
	\begin{itemize}
		\item Randomly choose a spin pulse.
		\item Calculate the gradient $g^{k}=\bigtriangledown _{I^{k}}Loss(I^{k})$.
		\item Calculate the exponential moving averages\\
		$m^{k}=\beta _{1}m^{k-1}+(1-\beta_{1})g^{k}$,\\
		$\upsilon^{k}=\beta _{2}\upsilon^{k-1}+(1-\beta_{2}(g^{k})^{2}$.
		\item Calculate the bias-corrected moment vectors.\\
		$\hat{m^{k}}=m^{k}/[1-(\beta_{1})^k],\hat{\upsilon^k}=\upsilon^k/[1-(\beta_{2})^k]$
		\item Update the pulse $I^k=I^{k-1}-\alpha\hat{m^{k}}/(\sqrt{\hat{\upsilon^k}}+\varepsilon)$.
		\item \textbf{Break} if $1-F(J^{N})<\xi$ or $k>k_{max}$.
	\end{itemize}
	\textbf{end for}
\end{algorithm}

\begin{figure*}
	\centering
	\subfigure{
		\begin{minipage}[t]{0.33\linewidth}
			\centering
			\includegraphics[width=2.3in]{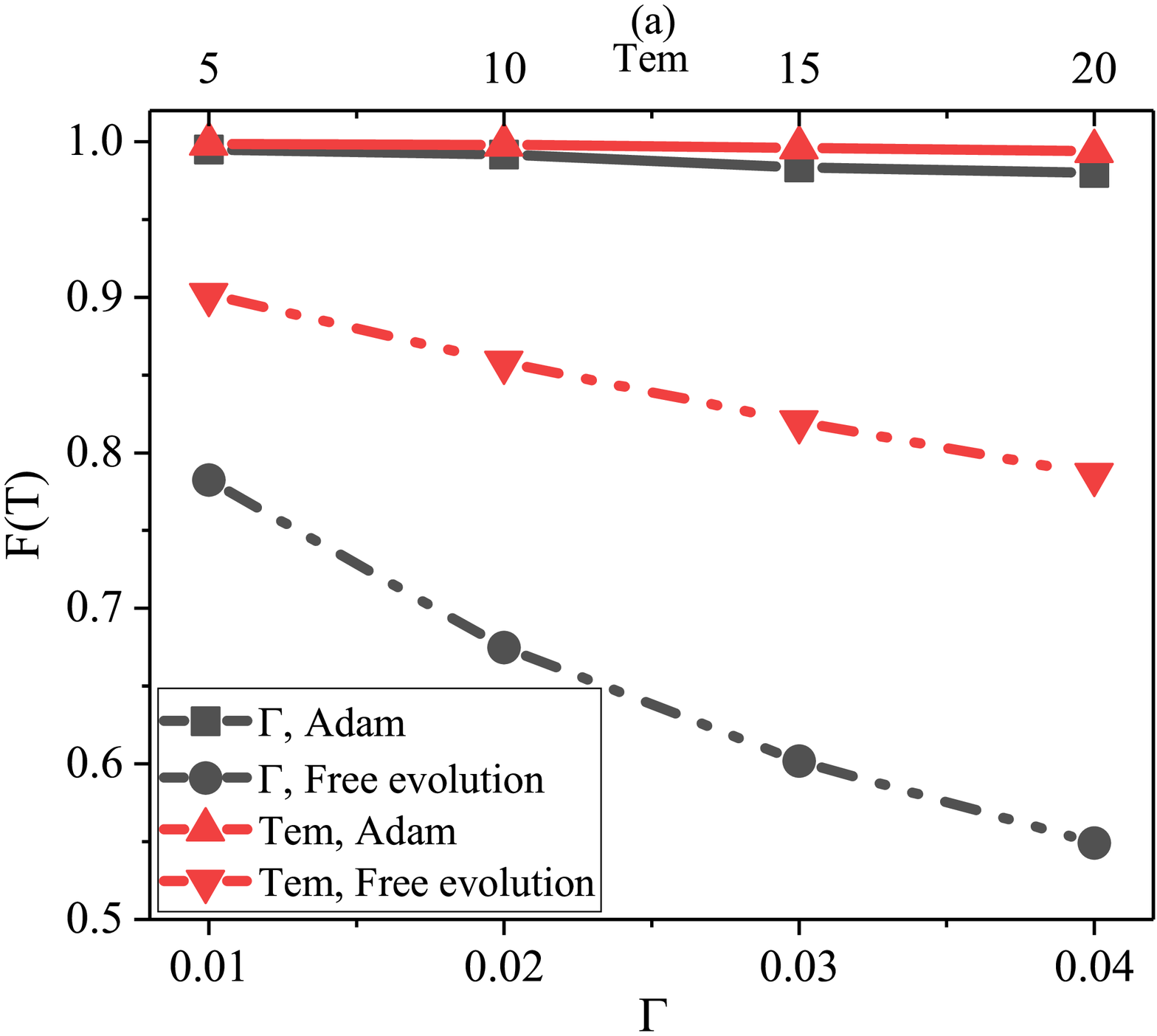}\\
		\end{minipage}%
	}%
	\subfigure{
		\begin{minipage}[t]{0.33\linewidth}
			\centering
			\includegraphics[width=2.3in]{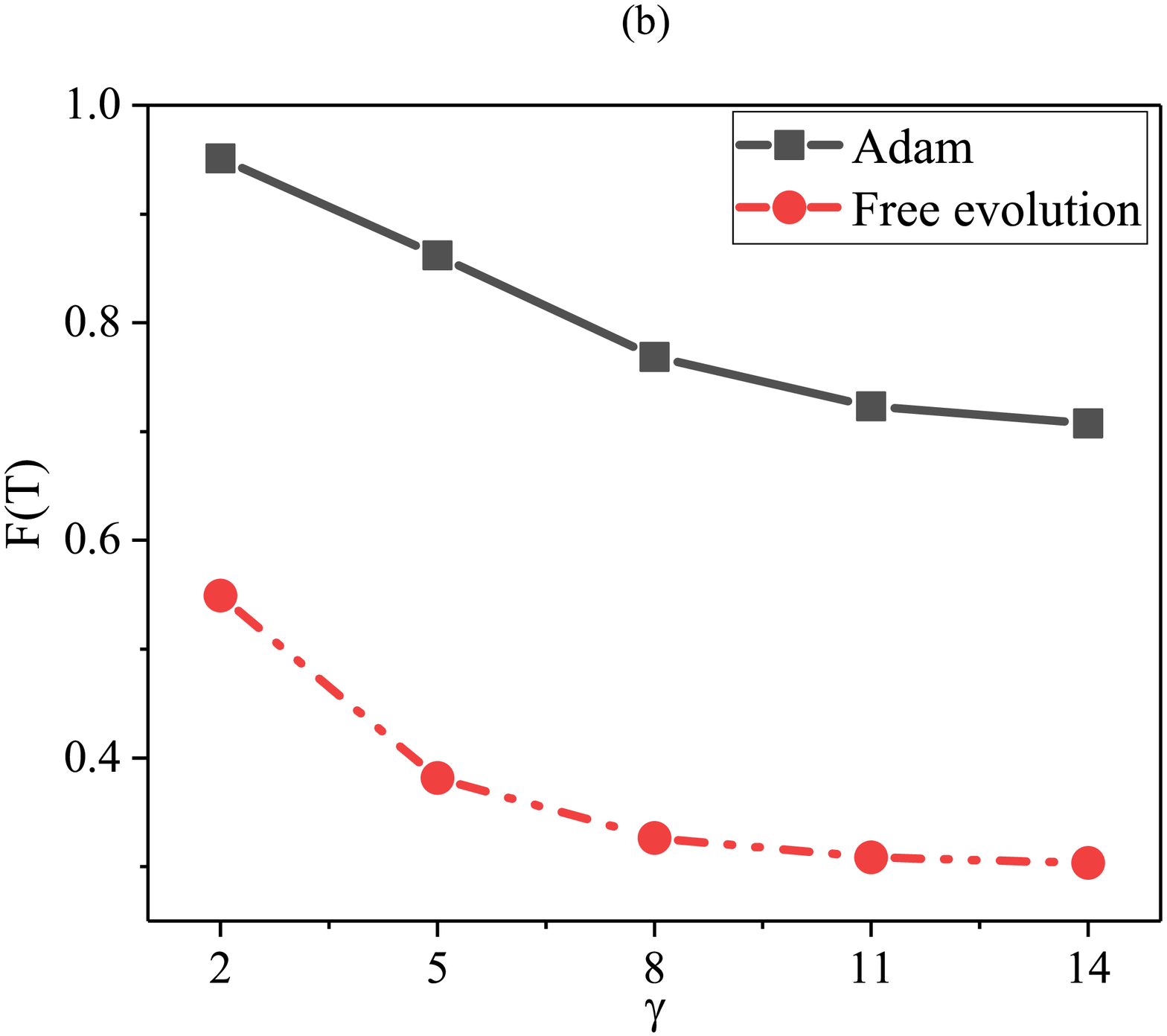}\\
		\end{minipage}%
	}%
	\subfigure{
		\begin{minipage}[t]{0.33\linewidth}
			\centering
			\includegraphics[width=2.3in]{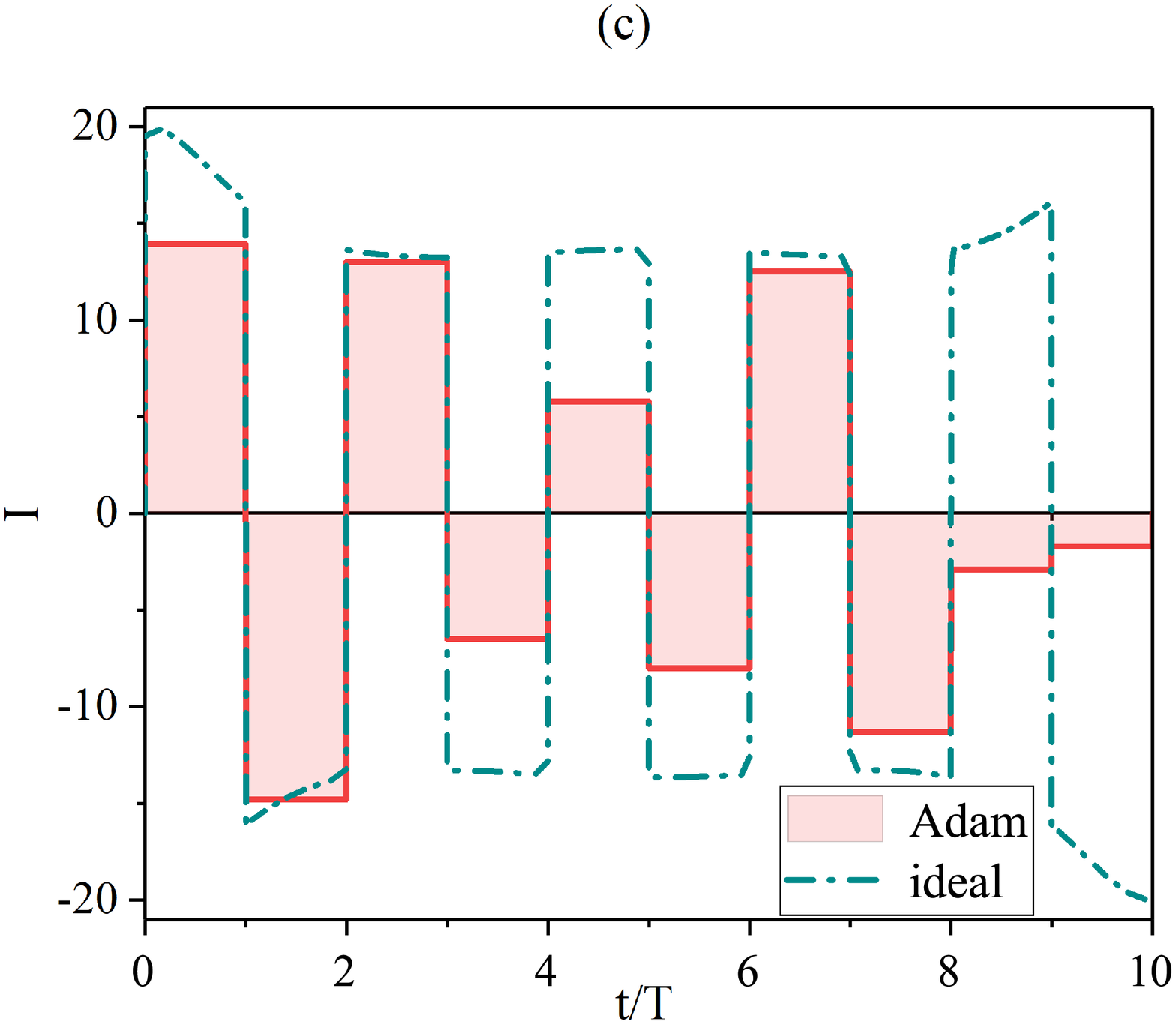}\\
		\end{minipage}%
	}%
	\centering
	\caption{The fidelity $F(T)$ with and without optimal pulses for different parameters. Only one environmental parameter is changed per line, the rest of the environmental parameters are the same. $L=\Sigma_{i}\sigma_i^-$. (a) $\gamma=2,Tem=10$ for $\Gamma$. $\Gamma=0.005,\gamma=2$ for $Tem$. (b) $\Gamma=0.04,Tem=10$ for $\gamma$. (c) The profile of the ideal pulses and Adam pulses. The environmental parameters $\Gamma=0.04,\gamma=14$ and $Tem=10$.}
	\label{fig:7}
\end{figure*}

The selection of the initial control pulses $I(t)$ is either by experience or guess. $\beta_{1}$ ($\beta_{2}$) is the decay exponent of the first (second) moment estimate. This method is computationally efficient and requires less memory. By updating $\boldsymbol{g_{t}},\boldsymbol{m_{t}},\boldsymbol{v_{t}}$, we can optimize the pulses to improve the fidelity. The iteration is terminated if the loss function $Loss(I^k)$ after the iteration is less than a given threshold $\xi$ $(\xi=0.001)$ or the iteration times $k>k_{max}$.  If the fidelity $F$ is improved, we keep the updated pulse $I^k$. Otherwise, we discard it. In this way, the pulse $I$ is gradually optimized and finally the optimal solution is obtained.

We first check the effectiveness of the Adam. In Fig.~\ref{fig:2} we plot the convergence behavior of the algorithm. In the optimization, we set the final fidelity $F$ to 0.999, correspondingly the parameter $\xi=0.001$. As an example, the environmental parameters are taken as $\Gamma=0.003, 0.005, 0.007$, $\gamma=2$, $Tem=10$, $T=\pi$, $N=5$, $\tau=\pi/10$. At this case the learning rate $\alpha$ in the Adam algorithm is chosen to be 1, and parameters $\beta_1=0.9$, $\beta_2=0.999$. 

From Fig.~\ref{fig:2} we see that the algorithm converges quickly: after about 3000 iteration the steady value is obtained. Therefore, the maximum number of iterations 3000 is chosen in this paper. We also find that  when $\Gamma$ becomes larger, the final fidelity $F(T)$ is smaller which will be discussed later.

\section{Results and Discussions}

In this section, we will use the Adam algorithm to design the zero-area pulses for high fidelity non-adiabatic QST. We will also compare the performances of the ideal pulses which are derived from the closed system \cite{zhang2019adiabatic,chen2018acceleration} and Adam optimized pulses. We use the rectangular pulses as in Eq.~(13) and define them as ideal pulses. For Adam optimized pulses, we take $I=10$ as our initial guess, which is different from the ideal pulses.  

At first, we analyze the effects of the environment on the transmission fidelity. In Fig.~\ref{fig:7}, we plot the final fidelity $F(T)$ via Adam optimization as a function of the environmental parameters $\Gamma$, $\gamma$, and $T$, respectively. To show the performance of the control, we also plot $F(T)$ without control. Clearly, the optimal pulses designed by the Adam shows its effectiveness: the near perfect QST can be realized even for a stronger bath (bigger $\Gamma$, $Tem$ and $\gamma$). From Fig.~\ref{fig:7}(b), $F(T)$ decreases with increasing $\gamma$, this is in accordance with previous results: a non-Markovian bath will be helpful to realize an effective transmission control that the fidelity can be boosted \cite{wang2020almost}. We also plot the pulse intensity $I(t)$ as a function of the rescaled time $t/T$ in Fig.~\ref{fig:7}(c). For the ideal pulses, $I(t)$ is tuned by the energy gap. Though the intensity of the Adam pulses are different in different period, it is a constant in half period. So the Adam pulses might be more easy to realize in the experiment.



\begin{figure}
	{\includegraphics[scale=0.33]{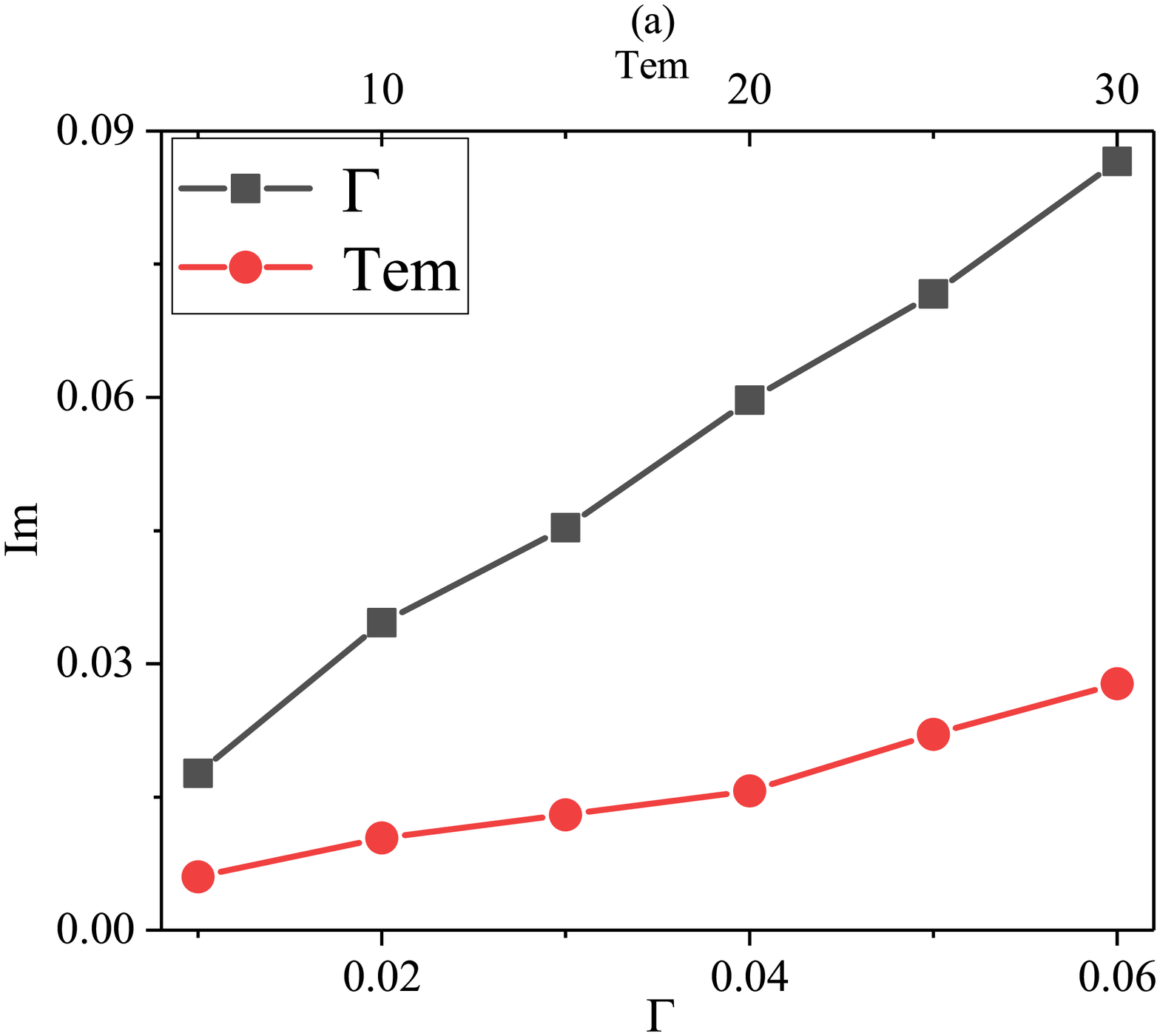}}
	{\includegraphics[scale=0.33]{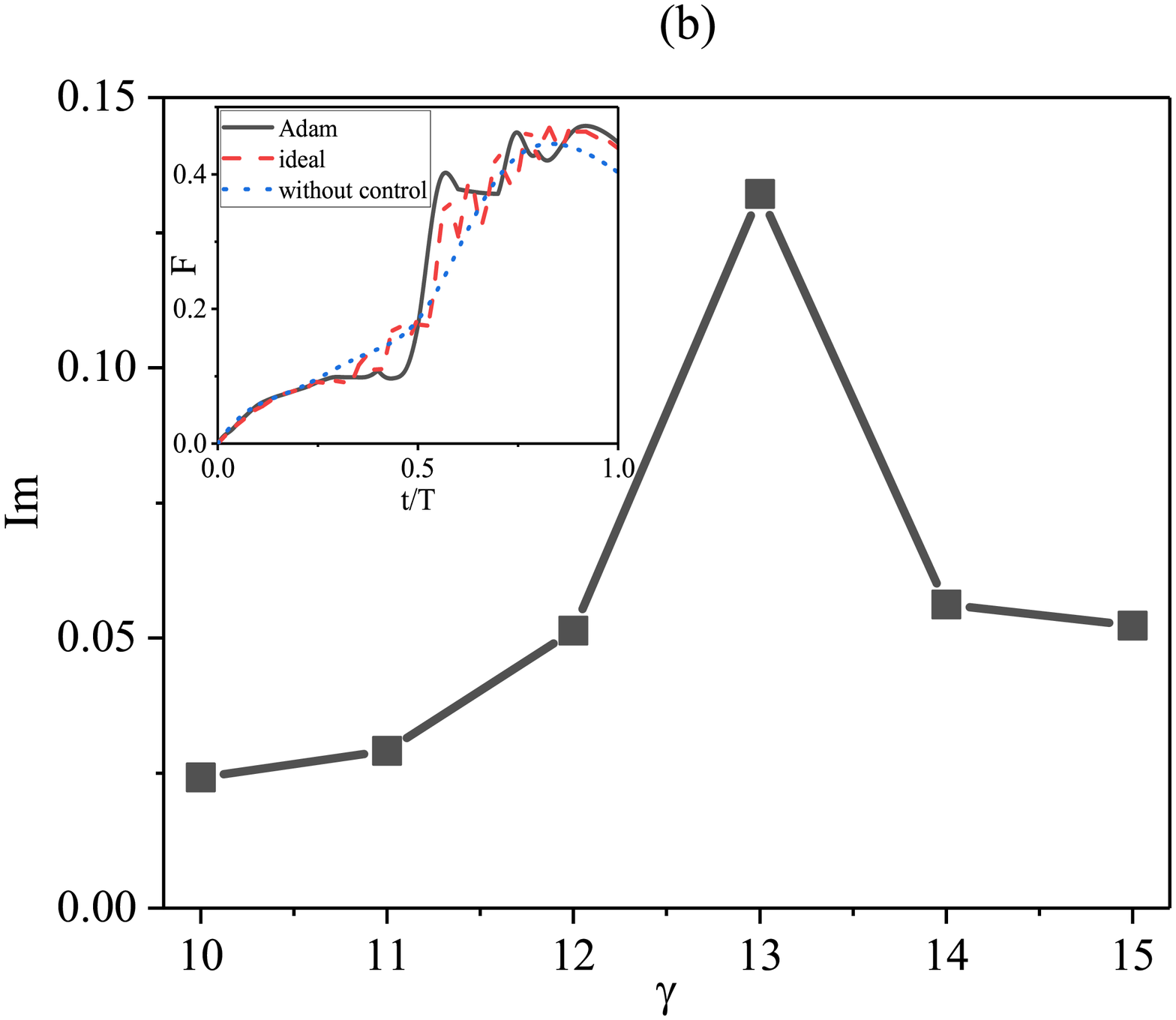}}
	\caption{The fidelity improvement $Im$ for different environmental parameters. $L=\Sigma_{i}\sigma_i^-$. (a) $\Gamma$ and $Tem$. For different $\Gamma$, $\gamma=2,Tem=10$. For different $Tem$, $\Gamma=0.005,\gamma=2$. (b) $\gamma$. $\Gamma=0.04,Tem=10$.}
	\label{fig:6}
\end{figure}

\begin{figure}
	\includegraphics[scale=0.33]{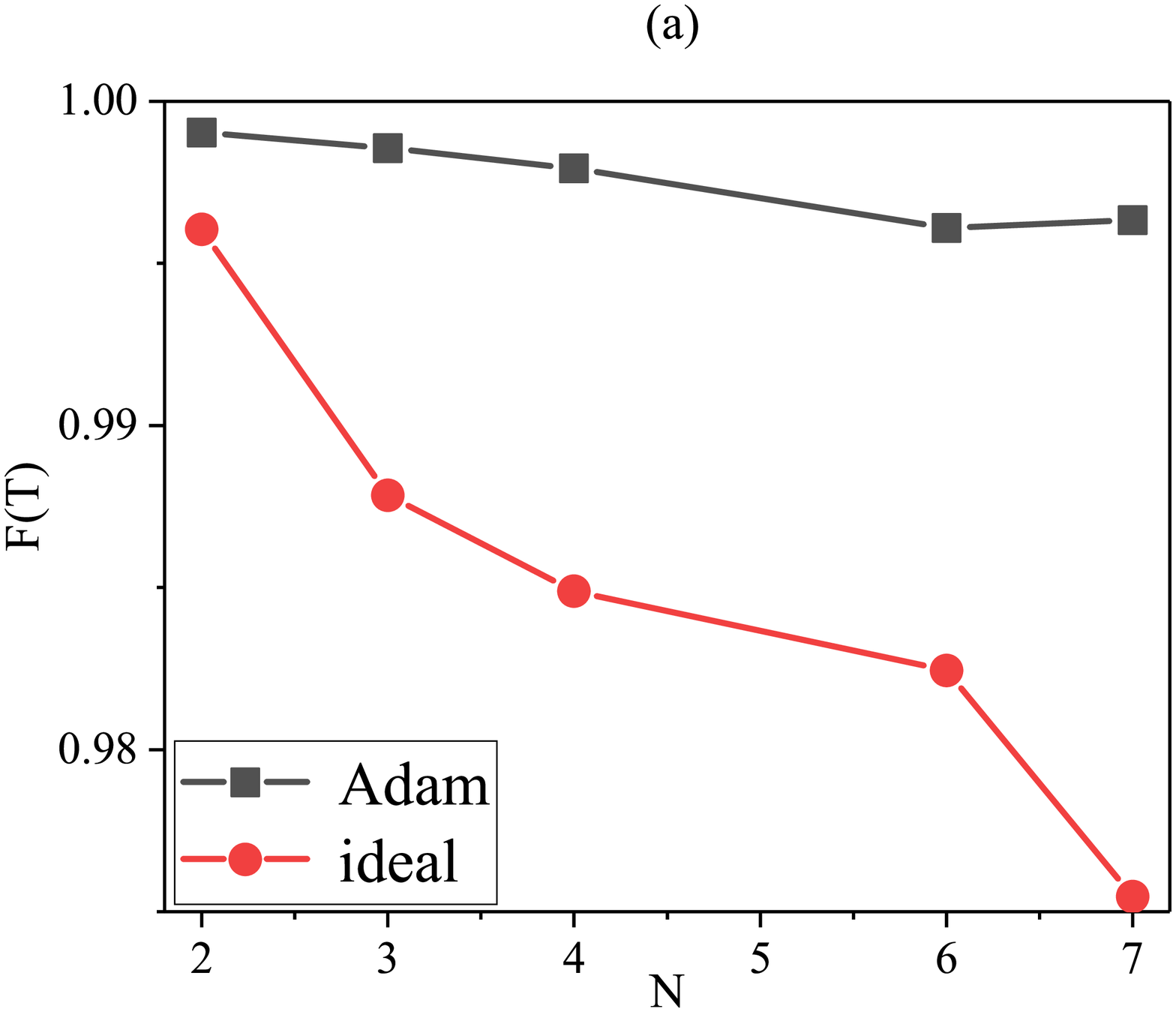}
	\includegraphics[scale=0.33]{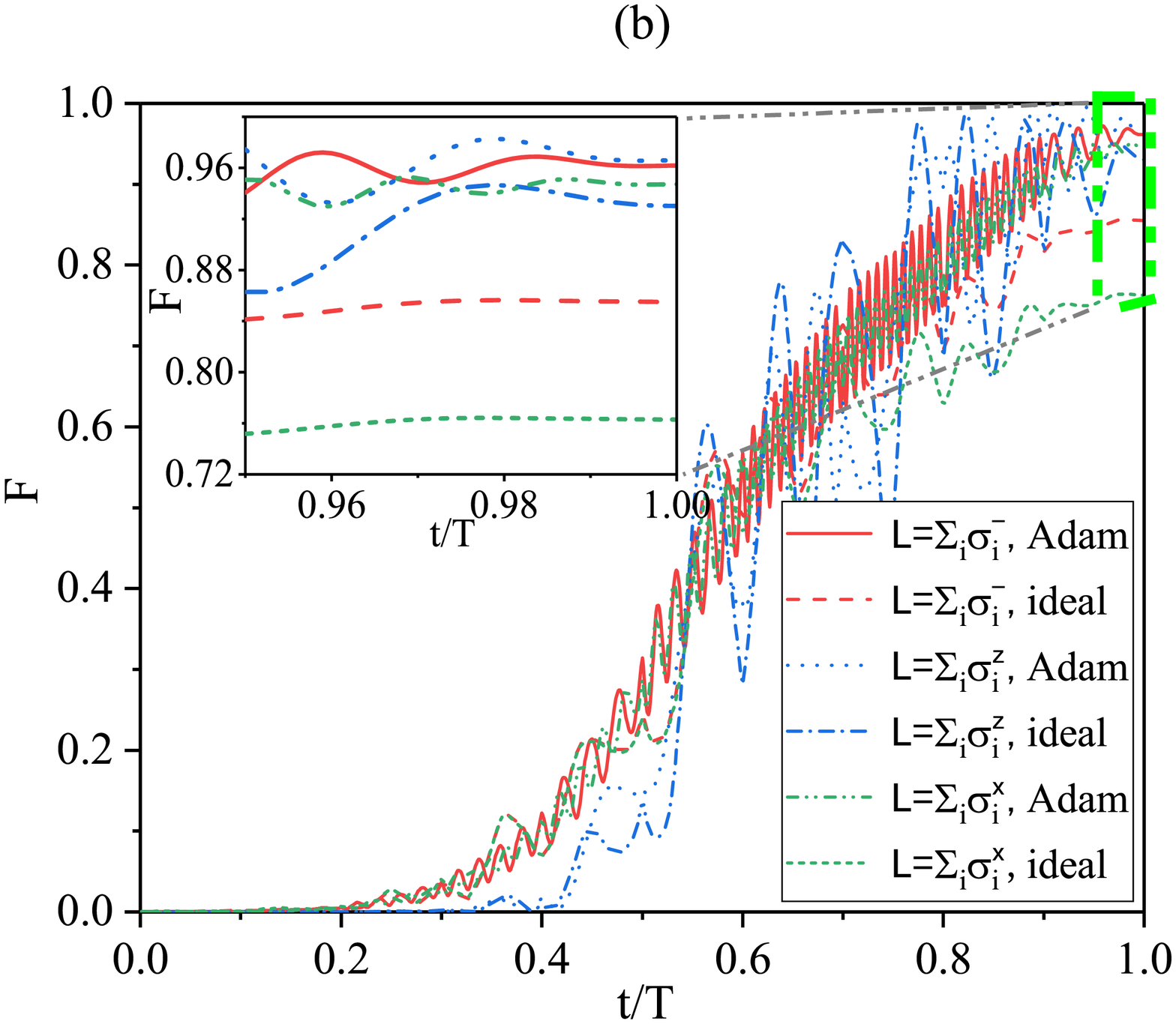}
	\caption{(a) Variation of the fidelity $F(T)$ vs the number of sites. Here $\Gamma=0.005,\gamma=2$ and $Tem=10$. (b) Change in fidelity $F$ vs time $t/T$ with the ideal pulses and the Adam pulses when $L=\Sigma_i\sigma_i^-,\Sigma_i\sigma_i^z,\Sigma_i\sigma_i^x$. $N=5,\Gamma=0.01,\gamma=8,Tem=15$.}
	\label{fig:8}
\end{figure}
To show the superiority of the Adam pulses to the ideal pulses, we calculate the fidelity improvement $Im$, which is defined by
\begin{equation}
	Im=F(T)^{Adam}-F(T)^{ideal}.
\end{equation}
where $F(T)^{Adam}$ and $F(T)^{ideal}$ are the final fidelity obtained from the Adam pulses and the ideal pulses, respectively.

Fig.~\ref{fig:6} plot the fidelity improvement $Im$ for different environemntal parameters $\Gamma$, $Tem$ and $\gamma$, respectively. For different $\Gamma$, $\gamma=2,Tem=10$. For different $\gamma$, $\Gamma=0.04,Tem=10$. For different $Tem$, $\Gamma=0.005,\gamma=2$. From Fig.~\ref{fig:6}(a), $Im$ increases with increasing $\Gamma$ or $Tem$. A stronger bath will destroy the system more and as a result the idea pulses lose its effectiveness because it is only valid in a closed system. Then the Adam algorithm shows its advantage in an open system.
Fig.~\ref{fig:6}(b) shows that $Im$ first increases then decreases with increasing $\gamma$. That is to say, for a more Markovian bath, the control loses its effectiveness for both Adam and ideal pulses, then $Im$ correspondingly becomes smaller.
The inset plot in Fig.~\ref{fig:6}(b) shows the variation of fidelity with time for the three cases in the Markov limit. The final fidelity after optimization of Adam algorithm is $F(T)$=0.4472. the fidelity under ideal pulse and without control are $F(T)$=0.4388 and $F(T)$=0.4029, respectively.
In the Markovian limit, the control is ineffective and $Im$ tends to be zero \cite{wang2020almost}. 
In sum, once the environmental parameters are ascertained, the corresponding pulses can be designed.

We only consider a fixed number of sites $N=5$ and the Lindblad operator $L=\Sigma_i\sigma_i^- $ in our previous discussion. Next we consider different $N$ and $L$. Fig.~\ref{fig:8}(a) plot $F(T)$ versus $N$ for ideal pulses and Adam pulses. As expected, $F(T)$ decreases with increasing $N$. However, the fidelity obtained by Adam pulses is always higher than the ideal case. For the Adam pulses, $F(T)$ decreases slowly with increasing $N$. Fig.~\ref{fig:8}(b) plots the time evolution of the fidelity $F$ for $L=\Sigma_i\sigma_i^-$, $\Sigma_i\sigma_i^x$ and $\Sigma_i\sigma_i^z$ with Adam and ideal pulses. The fidelity improvement of the Adam pulses can still be obtained for different $L$. In other words, the control scheme is still powerful. For $L=\Sigma_i\sigma_i^z$, the final fidelity $F(T)$ with Adam pulses is the biggest, $L=\Sigma_i\sigma_i^-$ is in the middle, and $L=\Sigma_i\sigma_i^x$ is the smallest.



\section{Conclusions}
Optimal control has been widely applied in different fields of physics. In this paper, we use the Adam algorithm, the extended version of stochastic gradient descent algorithm, to find the optimal pulses for the enhancement of the non-adiabatic QST fidelity in a non-Markovian environment. The model is a time-dependent one-dimensional spin chain in a finite-temperature heat bath. We use the non-Markovian quantum master equation, which is derived by the QSD technique, to calculate the dynamics of the chain. We find that the state transmission fidelity can be dramatically enhanced by the Adam pulses. Furthermore, we compare two kind of pulses: Adam pulses and ideal pulses. Though the fidelity can be enhanced by the ideal pulses, it is always lower than the Adam pulses because it is only valid in a closed system. The fidelity improvement $Im$ for these two cases ($\Gamma$ and $\gamma$) becomes larger for a more stronger bath, demonstrating the advantage of the Adam algorithm. Furthermore, we consider different length of the chain and types of the Lindblad operator. Our calculation results show that the Adam algorithm is still effective. Our investigation shows that the optimal control algorithm is a powerful tool to design pulses in performing quantum information processing tasks. 

\begin{acknowledgements}
	We would like to thank Ahmad Abliz, Yang-Yang Xie and Run-Hong He for their
	helpful discussions. This work was supported by the Natural Science Foundation of Shandong Province (Grant No. ZR2021LLZ004),  the Natural Science Foundation of China (Grant No. 11475160), the Spanish Grant No. PID2021- 126273NB- I00 funded by MCIN/AEI/10.13039/501100011033, and the Basque Government through Grant No. IT1470-22.
	\bibliographystyle{plainnat}
\bibliography{xianghan}
\end{acknowledgements}

\end{document}